\documentclass[12pt,preprint]{aastex}

\usepackage{epsfig}
\input psfig.sty 

 \newcommand{\eeq}{\end{equation}} 
\newcommand{\beq}{\begin{equation}}
\newcommand{\e}{\textrm{e}}

\def\la{\mathrel{\hbox{\rlap{\hbox{\lower4pt\hbox{$\sim$}}}\hbox{$<$}}}}
\def\sun{\hbox{$\odot$}}

\begin{document}
\title{On the dynamics of the universe in $D$ spatial dimensions}
\author{R. F. L. Holanda\altaffilmark{1}  }
\affil{Universidade Estadual da Para\'iba -- Departamento de F\'isica --
Rua Bara\'unas, 351 -- 58429-500 -- Bairro Universit\'ario -- Campina Grande, PB, Brazil}
\author{S. H. Pereira\altaffilmark{2}}
\affil{Universidade Federal de Itajub\'a -- Campus Itabira --
Rua Irm\~a Ivone Drumond, 200 -- 35903-087 -- Distrito Industrial II -- Itabira, MG, Brazil }

\altaffiltext{1}{holanda@uepb.edu.br}
\altaffiltext{2}{shpereira@gmail.com}

\begin{abstract}
In this paper we present the equations of the evolution of the universe in $D$ spatial dimensions, as a generalization of the work of Lima \citep{lima}. We discuss the Friedmann-Robertson-Walker cosmological equations in $D$ spatial dimensions for a simple fluid with equation of state $p=\omega_{D}\rho$. It is possible to reduce the multidimensional equations to the equation of a point particle system subject to a linear force. This force can be expressed as an oscillator equation, anti-oscillator  or a free particle equation, depending on the $k$ parameter of the spatial curvature. An interesting result is the independence on the dimension $D$ in a de Sitter evolution. We also stress the generality of this procedure with a cosmological $\Lambda$ term. A more interesting result is that the reduction of the dimensionality leads naturally to an accelerated expansion of the scale factor in the plane case. 
\end{abstract}

\section{Introduction}

The cosmological solutions for a relativistic simple fluid in the framework of Friedmann-Robertson-Walker (FRW) models were discussed long ago by Assad and Lima \citep{lima,AL88}. In the referred paper, the equation of cosmological dynamics driving the evolution of the scale factor for a simple perfect fluid obeying the equation of state $p=\omega\rho$, was reduced to the one of a point particle subject to a linear force, where $p$, $\rho$ and $\omega$ are the pressure, energy density and equation-of-state parameter describing the cosmic fluid, respectively. It has been demonstrated that the possible non-linear dynamic evolutions predicted by the FRW equations were naturally recovered in such reduction. In particular, closed models behave exactly as simple harmonic oscillators. The full discussion presented by Lima was restricted to 3+1 space-time dimensions \citep{lima}. Nevertheless, after the studies of Ehrenfest \citep{ehren}, who solved the Kepler problem in arbitrary dimensions, and Kaluza and Klein, in the 1920s, the interest for theories in $D$ spatial dimensions has grown considerably \citep{KK}. In 1991 Hayashi et. al. studied the influence of the dimension in physical laws \citep{hayashi}. More recently, cosmological models in higher dimensional space-time have been studied, called $D$-brane cosmology \citep{Dbrane,Dbrane2,Dbrane3,Dbrane4,Dbrane5,Dbrane6}. In this paper, we have explored the consequences of the method adopted by Lima for the study of a $D+1$ dimensional FRW cosmology, by studying the scale factor evolution as a function of $D$ for different material contents, e.g. matter, radiation and vacuum dominated universes. We study the generalization when a cosmological $\Lambda$-term is also present and we conclude by showing that the reduction of the dimensionality leads to an accelerated phase of expansion in the case of a plane ($k=0$) universe.

\section{FRW cosmologies in $D$ spatial dimensions}

The space-time metric for multidimensional FRW cosmologies in $D$ spatial dimensions is expressed as follows \citep{tan}, (we use $c=1$):
\begin{equation} 
\label{eq}ds^{2}=dt^{2}-a^{2}(t)\left(1+\frac{kr^{2}}{4}\right)^{-1}\delta_{ij}dx^{i}dx^{j} \,, \hspace{1cm} (i, j = 1,\, 2,\, 3 \,...\, D)
\end{equation} 
where $a(t)$ is the scale factor, $k$ is the curvature parameter of the spatial sections and $r^{2}=\sum_{i}(x_{i})^{2}$. The above expression reduces to the standard form in the 3-dimensional case.\citep{LL} 

In the $D$-dimensional geometry (\ref{eq}), Einstein's field equations for a relativistic simple fluid and the  energy conservation law can be written as:
\begin{equation} 
\label{eq1}\frac{D(D-1)}{2}\left[\frac{\dot a^{2}}{a^{2}}+\frac{k}{a^{2}}\right]=8\pi G_{D}\rho\,,
\end{equation} 
\begin{equation} 
\label{eq2}\frac{(D-1)\ddot a}{a}+ \frac{(D-1)(D-2)}{2}\left[\frac{\dot a^{2}}{a^{2}}+\frac{k}{a^{2}}\right]=-8\pi G_{D}p\,,
\end{equation} 

\begin{equation} 
\label{eq3}\dot \rho + D(\rho + p )\frac{\dot a}{a}=0\,,
\end{equation} 
where $G_D$, $\rho$ and $p$ are the $D$-dimensional Newtonian constant, the energy density and pressure of fluid, respectively. 

Following standard lines \citep{MW}, it will be assumed that the 
matter content obeys the general equation of state:
\begin{equation} 
\label{eq4} 
p=\omega_{D}\rho\,,
\end{equation} 
where $\omega_D$ is the equation-of-state parameter in $D$ spatial dimensions. For black-body radiation $\omega_D=1/D$, for matter $\omega_D=0$ and for vacuum $\omega_D=-1$. An interesting discussion on this equation regarding the adiabatic index $\gamma_D\equiv \omega_D+1$ can be found in the book  by Zeldovich and Novikov \citep{ZN96}.  For simplicity of notation, we will take $\omega_D \equiv \omega$.

By inserting the above expression into Eq. (\ref{eq3}) and integrating it, one may find that the energy density reads:     
\begin{equation} 
\label{eq5}\rho=\rho_{0}(\frac{a_{0}}{a})^{D(\omega+1)}\,,
\end{equation} 
where the cosmic scale factor, $a(t)$, must be determined from the FRW differential equation (see below). By combining Eqs. (\ref{eq1}), (\ref{eq2}) and (\ref{eq4}), it can be seen that the 
evolution of the scale factor is driven by the 
second order differential equation (which correctly reproduces the 3-dimensional case \citep{AL88,F99,LMJ98}): 
\begin{equation} 
\label{eq6}a \ddot a + \Delta_{D} \dot a^{2}+ \Delta_{D}k= 0\,,
\end{equation} 
where the definition 
\begin{equation} 
\Delta_{D}={D(\omega+1)\over 2}-1 \label{deltaD}
\end{equation} 
has been introduced. It is interesting to note that Eq. (\ref{eq6}) does not depend on the Newtonian constant $G_D$. Thus, we do not need to know its value in order to obtain the evolution.

In principle, the corresponding dynamic 
behavior must be heavily dependent on the 
choice of the following three free parameters: (i) The 
curvature parameter $k$, (ii) the equation of 
state parameter $\omega$, and (iii) the spatial dimension $D$. 

Now, let us discuss how the method of solution  proposed by Lima  in the 3-dimensional case can be extended for $D$ spatial dimensions \citep{lima}. This can be accomplished by using the conformal time $\eta$, instead of the cosmological or physical time, $dt=a(\eta)d\eta$. In this case, the equation of motion (\ref{eq6}) is expressed as shown below: 
\begin{equation} 
\label{eq7}a a^{''}+(\Delta_{D} -1)a^{'2}+\Delta_{D} ka^{2}=0 \,,
\end{equation} 
where a prime denotes differentiation with respect to conformal time $\eta$.

We now employ the auxiliary factor 
\begin{equation} 
\label{eq8}Z(\eta)=\ln a \ \ \mbox{if}\ \ \Delta_{D}=0 \,,
\end{equation} 
\begin{equation} 
\label{eq9}Z(\eta)=a^{\Delta_{D}} \ \ \mbox{if} \ \ \Delta_{D} \neq 0 \,,
\end{equation} 
to obtain, respectively, 
\begin{equation} 
\label{eq10}Z^{''}=0 \ \ \mbox{if}\ \ \Delta_{D}=0 \,,
\end{equation} 
\begin{equation} 
\label{eq11}Z^{''}+k\Delta_{D}^{2}Z=0 \ \ \mbox{if} \ \ \Delta_{D} \neq 0\,. 
\end{equation} 

As expected, although considering that we are treating  FRW cosmologies in $D$ spatial dimensions, the Equations (\ref{eq10}) and (\ref{eq11}) are reduced to those found by Lima for the 3-dimensional case \citep{lima}. Note also that  Eq. (\ref{eq11})  describes the classical motion of a particle subject to a linear force. This force can be restoring or repulsive depending only on the sign of the curvature parameter. The general solution of Eqs. (\ref{eq10}) and (\ref{eq11}) can be written as:
\beq
Z=b_0\eta+c_0 \ \ \mbox{if}\ \ \Delta_{D}=0 \,,\label{solD0}
\eeq
\beq
Z={z_0\over \sqrt{k}}\sin [\sqrt{k}\Delta_D(\eta +\delta)] \ \ \mbox{if} \ \ \Delta_{D} \neq 0 \label{sols}\,,
\eeq
with $b_0$, $c_0$, $z_0$ and $\delta$ integration constants. By choosing $\delta=0$ and determining $z_0=a_0^{\Delta_D}$, the scale factor evolution can be obtained as the solutions of (\ref{eq8}) and (\ref{eq9}):
\beq
a(\eta)=a_0 e^{b_0 \eta}\ \ \mbox{if}\ \ \Delta_{D}=0 \,,\label{solD0a}
\eeq
\beq
a(\eta)=a_0\bigg({\sin [\sqrt{k}\Delta_D\eta]\over \sqrt{k}}\bigg)^{1\over \Delta_D}  \ \ \mbox{if} \ \ \Delta_{D} \neq 0 \label{sols2}\,.
\eeq
The range in conformal time $\eta$ in flat ($k=0$) and open ($k=-1$) universes is semi-infinite, $+\infty > \eta > 0$, regardless of whether the universe is dominated by radiation ($\omega = 1/D$) or matter ($\omega = 0$). For a closed universe ($k=1$), $\eta$ is bounded to $\pi > \eta > 0$ for radiation and to $2\pi > \eta > 0$ for matter dominated universes \citep{mukh}.


For a flat space-time $(k=0)$  the system behaves like a free particle and the same happens if $\Delta_{D}= 0$, however, in the later case, this free particle behavior holds regardless of the curvature parameter. For $\Delta_D=0$ the Eq. (\ref{solD0a}) can be inverted by using $dt=a(\eta)d\eta$, and apart from integration constants, we obtain:
\beq
a(t)=b_0t\,.\label{abt0}
\eeq

For $\Delta_D \neq 0$, in the limit $k \to 0$, we have for (\ref{sols2}):
\beq
a(\eta)=a_0(\Delta_D\eta)^{1\over\Delta_D}\,.\label{s1}
\eeq
For $\Delta_D < 0$ the range of $\eta$ is $-\infty < \eta < 0$. In the specific case of flat universe the parametric solution can also be inverted to give the scale factor as a function of the cosmological time. Apart from an integration constant, we have:
\beq
a(t)=d_0\big(1+\Delta_D \big)^{1\over 1+\Delta_D} t^{1\over 1+\Delta_D}\,.\label{eqq18}
\eeq
It is easy to see that this expression reduces to (\ref{abt0}) in the limit $\Delta_D=0$ and identifying $b_0 = d_0$. We can see that, in the case $D=3$, for $\omega=0$ and $\omega =1/3$ we have the correct dependence $a\propto t^{2/3}$ and $a\propto t^{1/2}$ for matter- and radiation-dominated universes respectively. 

\begin{figure}[h!]
{\includegraphics[scale=0.8]{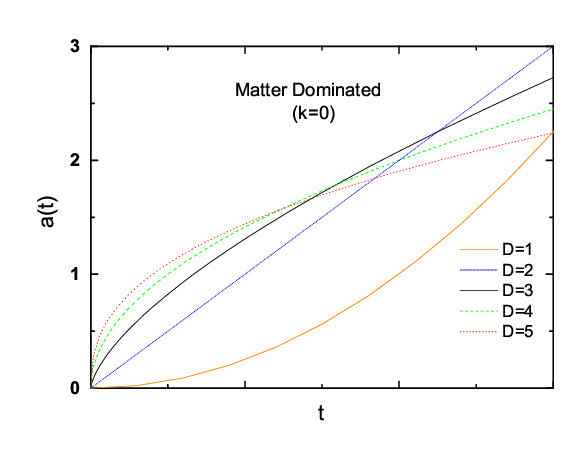}
\includegraphics[scale=0.8]{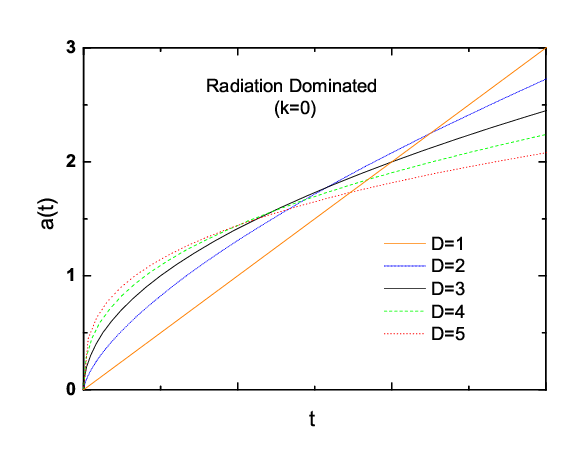}
\hskip 0.1in} \caption{ Scale factor evolution for different values of $D$ in the case of flat universe ($k=0$), for matter (top) and radiation (bottom) dominated universes. The horizontal time scale is arbitrary.} \label{fig1}
\end{figure}

In the Fig. \ref{fig1} we represent the scale factor evolution for different values of $D$ in the case of flat universe ($k=0$), for matter and radiation dominated universes. The expansion rate grows as the dimensionality increases. In the future, however,  the opposite behavior is observed, with reduction of the scale factor as $D$ increases.

To finish the study of the flat case, let us see how the age of the universe depends on the dimensionality. Deriving Eq. (\ref{eqq18}) with respect to time and taking $\dot a/a$, we obtain for the present time:
\beq
t_{0}={2\over D(\omega +1)}H^{-1}_{0}\,,
\eeq
where $H_0$ is the current Hubble parameter. This shows that the age of the universe decreases as dimensionality increases. Interestingly, if $D=3$ and $\omega=0$ (matter case) we recover $t_{0}=2/3H^{-1}_{0}$, and for $D=3$ and $\omega = 1/3$ (radiation case), we have $t_{0}=1/2H^{-1}_{0}$.


Closed models ($k=1$) are, for any value of $\Delta_{D}\neq 0$, analogous to simple harmonic oscillators. The cosmic dynamics in this case is similar to a spring-mass system where the spring constant is determined by the $\omega$-parameter and the number $D$ of spatial dimensions. The solution (\ref{sols2}) turns:
\beq
a(\eta)=a_0\big(\sin [\Delta_D\eta]\big)^{1\over \Delta_D}\,.\label{s2}
\eeq
Unlike the previous case, it is not possible to write directly the scale factor as function of the physical time $t$. 


For open space-times $(k=-1)$, the system behaves as a particle subject to a repulsive force proportional to the distance, or an anti-oscillator. The solution (\ref{sols2}) turns
\beq
a(\eta)=a_0\big(\sinh [\Delta_D\eta]\big)^{1\over \Delta_D}\,.\label{s3}
\eeq

In both cases, namelly $k=1$ and $k=-1$, numerical results shows that the evolution is always decelerating. 

\section{Vacuum ($\omega = -1$) dominated universe}

The case of a vacuum dominated universe ($\omega = -1$) is an interesting one. Note that the parameter $\Delta_D$ in (\ref{deltaD}) is independent of dimension in this case, $\Delta_D\equiv -1$. Consequently, the solutions (\ref{s1}), (\ref{s2}) and (\ref{s3}) are all independent of $D$. For the flat case for instance, we have:
\beq
a(\eta)=-{a_0\over \eta}\,\,,\hspace{1cm}-\infty<\eta<0\,,
\eeq
or
\beq
a(t)=a_0\e^{\alpha t}\,,
\eeq
with $\alpha$ a constant, which represents a de Sitter evolution of the scale factor.

\section{Evolution with a cosmological $\Lambda$ term}

As it happens  in the 3-dimensional case, we stress that the above method based on the transforming Equations (\ref{eq8}) and (\ref{eq9}) is also convenient when new ingredients are considered, such as the presence 
of a cosmological $\Lambda$ term. In this case, the second-order differential equation (\ref{eq6}) is as follows: 
\begin{equation} 
\label{eq12}a\ddot a + \Delta_{D}\dot a^{2}+ \Delta_{D}k=\frac{\Lambda (\omega+1) a^{2}}{D-1}\,. 
\end{equation}

One may show that the generalized equation of motion for  models with $\Lambda \neq  0$ in terms of the auxiliary scale factor is given by:
\begin{equation} 
\label{lam}Z^{''} + \Delta_{D}^{2}kZ=\frac{2\Delta_{D}(\Delta_{D} +1)\Lambda Z^{\frac{\Delta_{D} +2}{\Delta_{D}}}}{D(D-1)}\,.
\end{equation} 

The equation of motion (\ref{lam}) means that closed universes with 
cosmological constant evolve like anharmonic or non-linear oscillators. 
The anharmonic contribution to the oscillator is proportional to the cosmological 
$\Lambda$-term and inversely proportional to the number of space dimensions $D$. Its power index depends uniquely on the equation-of-state $\omega$-parameter. As expected, when $D \to 3$ the corresponding tridimensional result is  recovered (see Eq. (27) in Ref. \citep{lima}). 

In the particular case of a plane universe ($k=0$), an exact solution of (\ref{eq12}) can be obtained for arbitrary dimension $D$ and equation of state parameter $\omega$. The solution is
\beq
a(t)=\Bigg({D(D-1)\over 8\Lambda}\Bigg)^{1\over D(\omega+1)}{\Bigg[\exp\bigg((\omega +1)\sqrt{2\Lambda D\over D-1}t\bigg)-1\Bigg]^{2\over D(\omega+1)}\over \exp\sqrt{2\Lambda\over D(D-1)}t}\,,\label{solgen}
\eeq
where we have choosed the initial condition $a(0)=0$. In the case $D=3$ and $\Lambda \to 0$, this expression has the correct dependence $a\propto t^{2/3}$ and $a\propto t^{1/2}$
for $\omega=0$ and $\omega =1/3$ for matter and radiation-dominated universes, respectively. A graphycal analysis for different values of $D$ shows that the expansion starts decelerated but is always accelerated in the future for matter and radiation-dominated universe, as occur in the tridimensional case. As far as we know, the expression (\ref{solgen}) is presented here for the first time.

\section{Acceleration driven by the reduction of the dimensionality}

Another very interesting feature that we can also observe from Fig. 1 for the plane case is that the expansion of the scale factor is decelerated for large values of $D$, but clearly for some value of the dimension the expansion becomes accelerated, as indicated by the case $D=1$ in the matter dominated universe. In fact, in the case of matter ($\omega = 0$), the transition from decelerated to accelerated expansion occurs in $D = 2$, indicated by the linear expansion in the Figure. Thus, for $D> 2$ the expansion is decelerated, but for $D <2$ the expansion becomes accelerated. The same behavior occurs in the case of radiation dominated universe ($\omega = 1/3 $), where the transition occurs for $D = 1$. 

We realize that the reduction of the dimensionality can lead naturally to an accelerated expansion of the scale factor. Looking more closely to Eq. (\ref{eqq18}) we see that the transition from decelerated to accelerated phase occurs when the exponent of time becomes greater than 1. Thus for $ 1/(1 + \Delta_D) < 1$ the expansion is decelerated and for $ 1/(1 + \Delta_D)>1 $ the expansion is accelerated. The case $ 1/(1 + \Delta_D) = 1 $ represents the transition and can be written in terms of the equation of state parameter $ \omega $ as $ D = 2/ (\omega + 1) $. For $ \omega = 0 $ (matter), we obtain $ D = 2 $, and for $ \omega = 1/D $ (radiation) we have $ D = 1$, as already noticed before. An interesting consequence that follows is that for $ D =3$ the value of the equation of state parameter for which the transition from decelerated to accelerated stage occurs is $\omega = -1/3$. Such equation of state parameter characterizes the so called dark energy regime $\omega \leq -1/3$ \citep{lima3}. The range $\omega < -1$ represents the phantom regime \citep{sauloPhantom}, and it has been first suggested with basis on supernova analysis alone which favor $\omega < -1$ more than cosmological constant or quintessence \citep{cora}, and a more precise observational data analysis allows the equation of state parameter $\omega$ on the interval $[-1.38,\,-0,82]$ at 95\% confidence level \citep{melc}.

\section{Conclusion}

In this paper, we have studied the influence of the spatial dimension $D$ on the solutions of the FRW equations, as a generalization of the work of Lima \citep{lima}. We have shown that for both flat ($k=0$) and open ($k=-1$) universes, the increase in the dimensionality leads to a growth of the scale factor at the beginning of the evolution, but in the future the opposite behavior is observed, with reduction of the scale factor as $D$ increases. This occurs for both matter and radiation dominated universe.  For a closed universe ($k=1$) the behavior is that of a simple harmonic oscillator, with the collapse point shifted to small values of $t$ as $D$ increases. Generalized expression for the age of the universe in  $D$-dimensional spaces for a flat universe ($k=0$) was obtained, and we have shown that the increase in dimensionality implies a smaller value for the age of the  universe to reach the  actual size of the scale factor. This occurs for both matter and radiation dominated universe.

Another interesting conclusion is that the evolution for a vacuum dominated universe (de Sitter) does not depends on the spatial dimension $D$. The exacty and general solution that describes a universe in the presence of a cosmological $\Lambda$-term for arbitrary $D$ and $\omega$ is presented in the plane case ($k=0$), which correctly reproduces the corresponding tridimensional result for matter and radiation case.

A very interesting result is that the reduction of dimensionality leads naturally to an accelerated expansion. The dependence of $\Delta_D$ with the equation of state parameter $\omega$ shows that the transition from decelerated to accelerated regime occurs for $\omega < -1/3$, that characterizes a dark energy fluid. Such result is already known from other studies, but here it was obtained only with basis on the dimensional analysis.  This is a very important result, given the increasing number of studies in recent years involving the dynamics of the universe in extra dimensions.

\section*{Acknowledgments} 
RFLH is supported by CNPq, proc. num. 500201/2011-0 and SHP is supported by CNPq, proc. num. 477872/2010-7. We are very grateful to Professor Jos\'e A. S. Lima for suggesting this work and for many interesting comments.

\end{document}